\documentstyle[12pt]{article} 
\catcode`\@=11 
\def\marginnote#1{} 
\newcount\hour 
\newcount\minute 
\newtoks\amorpm 
\hour=\time\divide\hour by60 
\minute=\time{\multiply\hour by60 \global\advance\minute by-\hour} 
\edef\standardtime{{\ifnum\hour<12 \global\amorpm={am}%
        \else\global\amorpm={pm}\advance\hour by-12 \fi 
        \ifnum\hour=0 \hour=12 \fi 
        \number\hour:\ifnum\minute<10 0\fi\number\minute\the\amorpm}} 
\edef\militarytime{\number\hour:\ifnum\minute<10 0\fi\number\minute} 
\def\draftlabel#1{{\@bsphack\if@filesw {\let\thepage\relax 
   \xdef\@gtempa{\write\@auxout{\string 
      \newlabel{#1}{{\@currentlabel}{\thepage}}}}}\@gtempa 
   \if@nobreak \ifvmode\nobreak\fi\fi\fi\@esphack} 
        \gdef\@eqnlabel{#1}} 
\def\@eqnlabel{} 
\def\@vacuum{} 
\def\draftmarginnote#1{\marginpar{\raggedright\scriptsize\tt#1}} 
\def\draft{\oddsidemargin -.5truein 
        \def\@oddfoot{\sl preliminary draft \hfil 
        \rm\thepage\hfil\sl\today\quad\militarytime} 
        \let\@evenfoot\@oddfoot \overfullrule 3pt 
        \let\label=\draftlabel 
        \let\marginnote=\draftmarginnote 
   \def\@eqnnum{(\theequation)\rlap{\kern\marginparsep\tt\@eqnlabel}%
\global\let\@eqnlabel\@vacuum}  } 
 
\def\preprint{\twocolumn\sloppy\flushbottom\parindent 1em 
        \leftmargini 2em\leftmarginv .5em\leftmarginvi .5em 
        \oddsidemargin -.5in    \evensidemargin -.5in 
        \columnsep 15mm \footheight 0pt 
        \textwidth 250mmin      \topmargin  -.4in 
        \headheight 12pt \topskip .4in 
        \textheight 175mm 
        \footskip 0pt 
        \def\@oddhead{\thepage\hfil\addtocounter{page}{1}\thepage} 
        \let\@evenhead\@oddhead \def\@oddfoot{} \def\@evenfoot{} } 
 
\def\titlepage{\@restonecolfalse\if@twocolumn\@restonecoltrue\onecolumn 
     \else \newpage \fi \thispagestyle{empty}\c@page\z@  
        \def\thefootnote{\fnsymbol{footnote}} } 
 
\def\endtitlepage{\if@restonecol\twocolumn \else  \fi 
        \def\thefootnote{\arabic{footnote}} 
        \setcounter{footnote}{0}}  
 
\catcode`@=12 
\relax 
\def\beq{\begin{equation}} 
\def\eeq{\end{equation}} 
 
\def\NP#1#2#3{Nucl. Phys. \underline{#1} (19#2) #3} 
 
\def\PL#1#2#3{Phys. Lett. \underline{#1} (19#2) #3} 
\def\PR#1#2#3{Phys. Rev. \underline{#1} (19#2) #3} 
 
\def\Re{\mathop{\rm Re}}

\def\crbig{\\\noalign{\vspace {3mm}}}

\relax
 
\begin{document} 
\topmargin-2.4cm
\begin{titlepage} 
\begin{center} 
\hfill{NEIP--99--003} \\
\hfill{hep-th/9903064} \\
\hfill{March 1999} \\
\end{center} 
\vspace{1.1cm}
\begin{center}{\Large\bf 
Temperature Instabilities in N=4 Strings$^{\star}$ } 
\vskip .2in 
{\bf Jean-Pierre Derendinger$^{\dagger}$} 

\vspace{.6cm}
Physics Institute \\ 
Neuch\^atel University \\ 
CH--2000 Neuch\^atel, Switzerland 
\end{center} 

\vspace{1.1cm}
\begin{center} 
{\bf Abstract} 
\end{center} 
\begin{quote} 
An effective supergravity description of all instabilities 
of $N_4=4$ superstrings is derived. The construction is based on the
$N_4=4$ BPS mass formula at finite temperature and uses the properties of 
$N_4=4$ gauged supergravity. It provides the boundaries of the 
various thermal phases in the non-perturbative 
moduli space. It also draws a precise picture of the dynamics in
the high-temperature heterotic phase. 
This brief contribution summarizes results obtained in collaboration
with I. Antoniadis and C. Kounnas.
\end{quote} 

\vspace{2.1cm}
\begin{center}
{\it Invited contribution to the 32nd International Symposium Ahrenshoop 
on the Theory of Elementary Particles, Buckow, Germany, 
September 1--5, 1998.}
\end{center}
\vspace{3.5cm}

\begin{flushleft}
\rule{8.1cm}{0.2mm}\\
{\small
$^{\star}$
{\small Research supported by
the European Union under the TMR contract ERBFMRX-CT96-0045,
the Swiss National Science Foundation and the 
Swiss Office for Education and Science.} 
\\
$^\dagger$ jean-pierre.derendinger@iph.unine.ch }
\end{flushleft}

\end{titlepage} 
\setcounter{footnote}{0} 
\setcounter{page}{0}
\newpage 

\noindent
String theories (at weak coupling) generically exhibit high-temperature 
instabilities due to a density of states growing exponentially with the energy. 
At the Hagedorn temperature \cite{H}, a winding state becomes tachyonic 
and the string theory enters a new phase with a non-zero value of this state. 
The existence of the Hagedorn temperature can be established by studying 
the mass formula, derived from the modular 
invariant partition function at finite temperature \cite{T2}. Finite 
temperature is formally equivalent to a compactification 
of (euclidean) time on a circle with radius $(2\pi T)^{-1}$. 
Boundary conditions depending on statistics break supersymmetry 
and, because of modular invariance, modify GSO projections.

To obtain information on the phases in the vicinity of a Hagedorn transition, 
one approach is to construct an effective field theory for the light and 
tachyonic states. In Ref. \cite{AK}, this problem has been solved for
$N_4=4$ strings\footnote{$N_D$ is the number of $D$-dimensional 
supersymmetries.}. A $N_4=4$ supergravity is defined by a 
parametrization of its scalar manifold and a choice of the gauging 
applied to the vector fields present in the Yang-Mills and supergravity
multiplets. The gaugings related to, for instance, torus-compactified
heterotic strings \cite{GP} or
Scherk-Schwarz supersymmetry breaking \cite{PZ} are known. 
The knowledge of the specific gauging corresponding to finite temperature 
allowed the construction of effective supergravity Lagrangians
for the temperature instabilities of perturbative heterotic and type 
II strings and the study of the high-temperature phases \cite{AK}.

In five dimensions, heterotic strings on $T^4\times S^1$, IIA and IIB strings
on $K_3\times S^1$ are related by $S$- and $T$-dualities. 
A non-perturbative extension of the perturbative description 
of strings at finite temperature with a universal (duality-invariant) 
temperature modulus should then display an interesting
structure of thermal phases. These theories are effectively 
four-dimensional at finite temperature and an effective Lagrangian 
description of the thermal phases would be a four-dimensional supergravity. 
This construction has been performed in Ref.
\cite{ADK}\footnote{See also \cite{ADK2}.}, and 
the present contribution is a summary of this work.

Our procedure is firstly to write the finite-temperature generalization 
of the $N_4=4$ non-perturbative BPS mass formula, taking into account  
the expected dualities and the various perturbative heterotic, 
IIA and IIB limits. Secondly, an effective supergravity is constructed
by identifying the appropriate field content (potentially tachyonic states
and the minimal set of necessary moduli), the parametrization of the 
scalar manifold and, most importantly, the gauging for $N_4=4$ BPS states
at finite temperature. Notice that our analysis will be restricted to 
BPS states breaking half of the supersymmetries. 

In terms of the 
(heterotic) string coupling $g_H$ and the $T^2$ torus radii $R$ and $R_6$, 
the supersymmetric BPS mass formula is \cite{N=4BPS}:
\beq
\label{mass4}
\begin{array}{rcl}
{\cal M}^2 &=& \displaystyle{
\left[ {m\over R}+{nR\over\alpha_H^\prime}
+ g_H^{-2}\left({\tilde m^\prime\over R_6}+{\tilde n^\prime
R_6\over\alpha_H^\prime} \right) \right]^2
+\left[{m^\prime\over R_6}+{n^\prime R_6\over\alpha_H^\prime}
+ g_H^{-2}\left({\tilde m\over R}+{\tilde nR\over\alpha_H^\prime}
\right)\right]^2}
\crbig
&=& \displaystyle{{\left|
m+ntu +i (m^\prime u + n^\prime t) + is \left[
\tilde m + \tilde n tu - i(\tilde m^\prime u +\tilde n^\prime
t)\right]
\right|^2 \over \alpha_H^\prime tu}\,.}
\end{array}
\eeq
The integers $m,n,m^\prime, n^\prime$ are the
four electric momentum and winding numbers for the four
$U(1)$ charges from $T^2$ compactification. The numbers $\tilde m,
\tilde n, \tilde m^\prime, \tilde n^\prime$ are their magnetic
non-perturbative partners, from the heterotic point of view.
In the finite temperature case, 
the radius $R$ becomes the inverse temperature, 
$R=(2\pi T)^{-1}$ and the above mass formula is then modified to:
\beq
\label{mass5}
{\cal M}^2_T =
\left({m+Q'+{kp\over 2}\over R}+
k~T_{p,q,r}~R\right)^2-2 ~T_{p,q,r}~\delta_{|k|,1}
{}~\delta_{Q',0}\, ,
\eeq
where we have set $m'=n'={\tilde m}={\tilde n}=0$ to retain only 
the lightest states and $Q^\prime$ is the (space-time) helicity charge.
The integer $k$ is the common divisor of 
$(n,{\tilde m^\prime},{\tilde n^\prime})\equiv k(p,q,r)$
and $T_{p,q,r}$ is an effective string tension
$$
T_{p,q,r}={p\over\alpha_H^\prime}
+{q\over\lambda_H^2\alpha_H^\prime}
+{r R_6^2\over\lambda_H^2(\alpha_H^\prime)^2}\, ,
$$
with $\lambda_H^2 = g_H^2 RR_6/\alpha_H^\prime$ (six-dimensional
heterotic string coupling). 
The integer $\tilde m^\prime=kq$ is the wrapping number of
the heterotic five-brane around $T^4\times S^1_R$, 
while $\tilde n^\prime=kr$ corresponds to the same
wrapping number after performing a T-duality along the $S^1_{R_6}$
direction, which is orthogonal to the five-brane. All winding 
numbers $n, {\tilde m^\prime},
{\tilde n^\prime}$ are magnetic charges from the field
theory point of view. Their masses are proportional to the
temperature radius $R$ and are not thermally shifted.
A nicer writing of the effective string tension $T_{p,q,r}$ is
\beq
\label{Tpqr}
T_{p,q,r}={p\over\alpha_H^\prime}
+{q\over\alpha_{IIA}^\prime}
+{r\over\alpha_{IIB}^\prime}\,,
\eeq
where $\alpha_H^\prime = 2\kappa^2 s$, 
$\alpha_{IIA}^\prime = 2\kappa^2 t$ and  
$\alpha_{IIB}^\prime = 2\kappa^2 u$ 
when expressed in Planck units. 

The mass formula (\ref{mass5}) possesses the same duality properties
as the zero-temperature expression (\ref{mass4}) and $R$,
the inverse temperature, is a duality-invariant quantity.
Eq. (\ref{mass5}) gives the states and 
critical values of the temperature
radius at which a tachyon appears. Each corresponds to the 
Hagedorn transition of a perturbative string, either heterotic, 
or IIA or IIB. It also contains new information (on critical 
values of $\lambda_H$ and/or $R_6$) since it also
decides which tachyon arises first when $T\sim 1/R$ increases. 

We now want to construct an effective (four-dimensional) supergravity
for the five-dimensional strings at finite temperature. 
To describe instabilities, we may truncate the $N_4=4$
spectrum and retain only the necessary moduli and potentially tachyonic
states, as indicated by mass formula (\ref{mass5}). 
The resulting truncated theory will have $N_4=1$ supersymmetry\footnote{
The four gravitinos are treated identically by finite temperature effects.} 
and include chiral multiplets only.

The scalar manifold of a generic, {\it unbroken}, $N_4=4$ theory is
\cite{DF, dR}
\beq
\label{manif1}
\left({Sl(2,R) \over U(1)}\right)_S \times \, G/H, \qquad\qquad
G/H =
\left({SO(6,r+n)\over SO(6)\times SO(r+n)}\right)_{T_I,\phi_A}.
\eeq
The manifold $G/H$ of the vector multiplets
splits into a part that includes the $6r$
moduli $T_I$, and a second part with the infinite number
$n\rightarrow\infty$ of BPS states. We need to keep three moduli 
$S$, $T$ and $U$ (for the temperature radius $R$, the torus radius 
$R_6$ and the string coupling) and three pairs of winding states 
$Z_A^\pm$, $A=1,2,3$, to generate the instabilities, as indicated by 
mass formula (\ref{mass5}). Thus, $r=2$ and $n=6$ in Eq. (\ref{manif1}). 
The truncation from $N_4=4$ to $N_4=1$ proceeds then as for
the untwisted sector of a $Z_2\times Z_2$ orbifold. The
first $Z_2$ leaves $N_4=2$ unbroken and the manifold is
$$
\begin{array}{l}
\displaystyle{\left({Sl(2,R) \over U(1)}\right)_S \times
\left({Sl(2,R) \over U(1)}\right)_T \times
\left({Sl(2,R) \over U(1)}\right)_U \times
\left({SO(4,6)\over SO(4)\times SO(6)}\right)_{\phi_A}.}
\end{array}
$$
The first three factors are vector multiplets ($S$, $T$, $U$), the last
one a hypermultiplet component ($Z_A^\pm$). The second $Z_2$ truncation 
cuts the hypermultiplet component in two K\"ahler manifolds:
$$
\left({SO(2,3)\over SO(2)\times SO(3)}\right)_{Z_A^+}\times
\left({SO(2,3)\over SO(2)\times SO(3)}\right)_{Z_A^-}.
$$
The structure of the truncated scalar manifold and the Poincar\'e $N_4=4$
constraints on the scalar fields \cite{dR} indicate that the
K\"ahler potential can be written as \cite{AK, PZ}
\beq
\label{Kis}
K = -\log(S+S^*)-\log(T+T^*)-\log(U+U^*) 
-\log Y(Z_A^+,Z_A^{+*}) -\log Y(Z_A^-,Z_A^{-*}),
\eeq
with $Y(Z_A^\pm,Z_A^{\pm*}) = 1 -2Z_A^\pm Z_A^{\pm*} + (Z_A^\pm
Z_A^\pm)(Z_B^{\pm*}Z_B^{\pm*})$.
This K\"ahler function can be determined for instance 
by comparing the gravitino mass terms in the
$N_4=1$ Lagrangian and in the $Z_2\times
Z_2$ truncation of $N_4=4$ supergravity.

The last piece to define the effective supergravity is the 
superpotential. Finding the correct gauging of $N_4=4$ supergravity 
for mass formula (\ref{mass5}) requires some guesswork. We find:
\beq
\label{Wis}
W = 2\sqrt2 \left[ {1\over2}(1-Z_A^+Z_A^+)(1-Z_B^-Z_B^-)
+\,(TU-1)Z_1^+Z_1^- + SUZ_2^+Z_2^- +
STZ_3^+Z_3^-  \right].
\eeq
Eqs. (\ref{Kis}) and (\ref{Wis}) define an effective supergravity which 
includes in its solutions the thermal phases of five-dimensional 
$N_4=4$ heterotic, type IIA and type IIB strings, and respects all expected 
duality symmetries. 

The scalar potential is complicated but a closed expression can be worked 
out\footnote{The K\"ahler metric can be explicitly inverted.}. It is 
stationary at $Z_A^\pm=0$ (zero winding background value). At this point,
masses follow the formula (\ref{mass5}), as they should. This solution 
corresponds to the low-temperature phase. Tachyons only arise in 
directions $\Re(Z_A^++Z_A^-)\equiv z_A$, and truncating further to these
directions leads to a very simple potential:
\beq
\label{pot4}
\begin{array}{rcl}
V &=& V_1 + V_2 + V_3,  \crbig
\kappa^4 V_1 &=& \displaystyle{4\over s}\left[
(tu+{1\over tu})H_1^4
+{1\over4}(tu-6+{1\over tu})H_1^2 \right],
\crbig
\kappa^4 V_2 &=& \displaystyle{4\over t}
\left[ suH_2^4 +{1\over4}(su-4)H_2^2\right],
\crbig
\kappa^4 V_3 &=& \displaystyle{4\over u}
\left[ stH_3^4 +{1\over4}(st-4)H_3^2\right].
\end{array}
\eeq
In these expressions, $H_A=z_A/(1-z_Bz_B)$, $s=\Re S$, 
$t=\Re T$, $u=\Re U$, heterotic temperature duality is $tu\rightarrow
(tu)^{-1}$ and IIA--IIB duality is $t,H_2\leftrightarrow u,H_3$.

Limited space permits us to only briefly review 
the analysis of the effective theory.

\medskip
\noindent {\bf 1)} 
In four-dimensional Planck units, the temperature is duality-invariant:
\beq
\label{Tis}
T=(2\pi R)^{-1}, \qquad\qquad R^2 = \kappa^2 stu. 
\eeq

\medskip
\noindent {\bf 2)} 
The low-temperature phase, with zero background values of the
winding states $Z_A^\pm$, is universal to the three strings. It is in 
some sense a self-dual phase, as can be seen for instance in the
pattern of supersymmetry breaking. Since
\beq
\label{lowTsusy}
({\cal G}^S_S)^{-1/2}\,{\cal G}_S =
({\cal G}^T_T)^{-1/2}\,{\cal G}_T =
({\cal G}^U_U)^{-1/2}\,{\cal G}_U = -1
\eeq
(other derivatives of ${\cal G} = K+\log|W|^2$ vanish), 
the canonically normalized Goldstino is the sum of the fermionic partners
of $S$, $T$ and $U$. And the gravitino mass is
\beq
\label{mgavapp}
m_{3/2}^2 = \kappa^{-2}e^{\cal G} = {1\over4\kappa^2stu}
= {1\over4 R^2} = (\pi T)^2
= {1\over2\alpha^\prime_H tu} = {1\over2\alpha^\prime_{IIA} su}
= {1\over2\alpha^\prime_{IIB} st}.
\eeq
This phase certainly exists in the perturbative regime of each string. 

\medskip
\noindent {\bf 3)}
The boundaries of the low-temperature phase are the values 
of $s$, $t$, and $u$ at which a winding state becomes tachyonic:
\beq
\label{tachyons}
\begin{array}{rrcl}
{\rm heterotic \,\, tachyon}&\quad\Re(Z_1^++Z_1^-)\qquad&{\rm if}& \qquad
(\sqrt2+1)^2 > tu > (\sqrt2-1)^2,
\crbig 
{\rm type \,\, IIA \,\, tachyon}&\quad\Re(Z_2^++Z_2^-) \qquad&{\rm if}&
\qquad su < 4, \crbig
{\rm type \,\, IIB \,\, tachyon}&\quad\Re(Z_3^++Z_3^-) \qquad&{\rm if}&
\qquad st < 4.
\end{array}
\eeq
The boundaries are then $tu=(\sqrt2+1)^2$,
$su=4$ and $st=4$, or, in heterotic variables,
\beq
\label{bound}
R =\sqrt{\alpha_H^\prime}\,{\sqrt2+1\over\sqrt2}, \qquad
R = 2g_H^2 R_6, \qquad
R = 2g_H^2 {\alpha^\prime_H\over R_6} = {4\sqrt2\kappa^2\over R_6},
\eeq
with $\alpha_H^\prime=2\kappa^2s$.
At these values, $T$ is a Hagedorn temperature. 

\medskip
\noindent {\bf 4)}
Type II instabilities arise when $su<4$ (IIA) or $st<4$ (IIB).
From the heterotic point of view, they are avoided as long as
\beq
\label{Tineq}
2\pi T <  {1\over 4\sqrt2\kappa^2}\, {\rm min}\left( R_6\,\,;\,\,
\alpha^\prime_H/R_6\right).
\eeq
And type II instabilities are unavoidable if
\beq
\label{TallII}
2\pi T > {2^{1/4}\over 4\kappa g_H}.
\eeq

\medskip
\noindent {\bf 5)} 
In the high-temperature heterotic phase, 
$(\sqrt2+1)^2 > tu > (\sqrt2-1)^2$ and $su>4<st$.
It cannot be reached\footnote{From
low temperature.} for any value of the radius $R_6$ if
the (lowest) heterotic Hagedorn temperature verifies inequality 
(\ref{TallII}), which translates into
\beq
\label{glim}
g_H^2 > g^2_{\rm crit.}= {\sqrt2+1\over2\sqrt2} \,\sim \,0.8536.
\eeq
Only type
II thermal instabilities exist in this heterotic 
strong-coupling regime and the
value of $R_6/\sqrt{\alpha^\prime_H}$ decides whether the type IIA or
IIB instability will have the lowest critical temperature.
If on the contrary the heterotic string is weakly coupled,
$ g_H<g_{\rm crit.}$,
the high-temperature heterotic phase is reached for values of the
radius $R_6$ verifying
\beq
\label{R6ineq}
2\sqrt2g_H^2(\sqrt2-1) < {R_6\over\sqrt{\alpha^\prime_H}} <
{1\over2\sqrt2g_H^2(\sqrt2-1)}.
\eeq
The large and small $R_6$ limits, with fixed $g_H$, again
lead to type IIA or IIB instabilities.

\medskip
\noindent {\bf 6)}
In the high-temperature heterotic phase, after solving for 
$Z_A^\pm$, the potential becomes
$$
\kappa^4 V = -{1\over s}{(tu+{1\over tu}-6)^2\over
16(tu+{1\over tu})}.
$$
It has a stable minimum for fixed $s$ (for fixed $\alpha^\prime_H$).
In units of $\alpha^\prime_H$, the temperature is fixed,
$tu=1=2R^2/\alpha^\prime_H = R^2/(\kappa^2 s)$.
The transition from the low-temperature vacuum is due to a
condensation of the heterotic winding mode $\Re(Z_1^++Z_1^-)$, 
or equivalently by a condensation of NS five-brane in the type
IIA picture.

At $tu=1$, $\kappa^4V= -1/(2s)$ and the heterotic dilaton $s$ runs away.
The effective supergravity is solved by a background with a dilaton 
linear in a space-like direction. The fate of 
supersymmetry is interesting.
Inequality (\ref{glim}) indicates that the high-T heterotic phase 
only exists for weakly-coupled heterotic strings, 
and by duality in non-perturbative type II regimes.
Accordingly, supersymmetry breaking arises from $s$: only ${\cal G}_S$ 
is non-zero. It turns out, as observed in Ref. \cite{AK}, that
the spectrum of moduli and heterotic perturbative winding modes $Z_1^\pm$
is supersymmetric\footnote{Boson-fermion mass degeneracy.}.
But this degeneracy does not exist for heterotic dyonic modes
$Z_2^\pm$ and $Z_3^\pm$:
$$
\begin{array}{rrcl}
Z_2^\pm:&  m_{bosons}^2 &=& m_{fermions}^2 \pm 2su\, m_{3/2}^2,
\crbig
Z_3^\pm:&  m_{bosons}^2 &=& m_{fermions}^2 \pm 2st\, m_{3/2}^2.
\end{array}
$$
The significance of this mass pattern is somewhat ambiguous.
The residual space-time symmetry with a linear dilaton 
is three-dimensional and local supersymmetry does not imply mass degeneracy 
in this dimension \cite{3d}. 
This mass pattern however survives in five-dimensional type IIA and 
IIB limits, indicating then broken supersymmetry. 

\medskip
\noindent {\bf 7)}
The linear dilaton background leads to a central charge 
deficit\footnote{Superstring counting: $\delta\hat c = {3\over2}\delta c$.}
$\delta\hat c=-4$. It can be described by a non-critical string with the 
corresponding central charge. The appropriate string background is a 
non-compact parafermionic space, in which degeneracy of perturbative
bosonic and fermionic fluctuations is ensured by $N_4=2$ (or $N_3=4$)
supersymmetry.
This agrees with the space-time solution with linear dilaton:
the background is left invariant by the half of the supercharges in 
the $N_4=4$ algebra.

\vspace{8mm}
\noindent
{\large \bf Acknowledgements}

\medskip
\noindent
I wish to thank the
organizers of the 32nd International Symposium Ahrenshoop on 
the Theory of Elementary Particles. This work is supported by 
the European Union (contracts TMR-ERBFMRX-CT96-0045 and -0090), 
the Swiss Office for Education and Science and the Swiss National 
Science Foundation.

\end{document}